\documentstyle{article}

\input epsf
 \def\calV{{\cal V}}
\def\calS{{\cal S}}
\def\calT{{\cal T}}
\def\calU{{\cal U}}
\def\calW{{\cal W}}
\def\cals{{\cal S}_s}
\def\calt{{\cal T}_s}
\def\calu{{\cal U}_s}
\def\calw{{\cal W}_s}
\def\calR{{\cal R}}
\def\Rfour{t_8t_8R^4}
\def\lam16{\lambda^{16}}
\def\hl{\hat l}

\def\half{{\textstyle {1 \over 2}}}

\def\be{\begin{equation}}
\def\ee{\end{equation}}
\def\bes{\begin{eqnarray}}
\def\ees{\end{eqnarray}}
\def\nn{\nonumber}

\relax
\begin{document}

\thispagestyle{empty}

\null
\vskip-2.0cm
{\hfill DAMTP/97-50}
\vskip 0.1cm
{\hfill hep-th/9712195}
\vskip 1.5 cm

\begin{center}
{\large\bf Connections between M-theory and superstrings }
\vskip 1.0cm
 \centerline{Michael B.  Green}
\medskip
 {Department of Applied Mathematics and
Theoretical Physics, Silver Street, Cambridge CB3 9EW, United Kingdom}
 
\medskip
\centerline{\it M.B.Green@damtp.cam.ac.uk}
\vskip 0.3cm
 
\end{center}
\vskip 3.0 cm
\rm

\begin{quote}
{\bf ABSTRACT:}

This article reviews the non-perturbative  structure  of  certain higher
derivative  terms in  the type II string theory effective action  and their
connection to  one-loop effects in eleven-dimensional supergravity compactified
on a  torus.  New material is also included that was not presented in the talks.
\footnote{Based on talks given at the ICTP Conference on Duality Symmetries in String Theory, 7-12 April 1997; STRINGS '97, Amsterdam, June 16-21 1997;    CERN Workshop on Non-perturbative Aspects of Strings, Branes and Fields,  8-12 December 1997. }

\end{quote}
\vskip1cm

\normalsize

\newpage
\pagestyle{plain}
\setcounter{page}{1}

\def\baselinestretch{1.2}
\baselineskip 16 pt
\noindent
 \setcounter{equation}{0}

\section{Introduction}

Many features of M-theory follow directly from the strong algebraic constraints
associated with the symmetries of the theory  rather than specific details of
any microscopic model, such as the matrix model.   Certain  of these
properties  must be inevitable consequences of  eleven-dimensional
supersymmetry.   Others may depend on assuming the validity of the intricate
web of duality inter-relationships between the perturbative string theories.
 Presumably a microscopic description, perhaps  based on the matrix model,  is needed to define the theory on all length scales  --- indeed, to define the very notion of a length scale.
But it is of interest to determine  the  extent to which   terms in
 a low-momentum expansion of  the M-theory  effective action are  determined simply by
eleven-dimensional supergravity compactified in various ways.  

 It has been
generally assumed that since perturbative  supergravity is terribly divergent in eleven
dimensions no interesting consequences follow by considering its  loop diagrams.
However, this ignores the powerful  supersymmetry constraints   which are
expected to reduce the number of arbitrary constants -- but by how much?  Furthermore, the eleven-dimensional quantum theory compactified to ten or fewer
dimensions  is supposed to be precisely equivalent to string theory and its
compactifications.   What mileage can be gained by exploiting this equivalence
as an   ansatz? The lowest order terms in the expansion of  the M-theory action
in powers of the momentum are the terms in the original classical supergravity
action of \cite{cjs}.    Certain higher order terms  have also been
unambigously identified.   One example   is the $C^{(3)} \wedge X_8$ term
\cite{minasduff} that arises as a one-loop effect in type IIA string theory
\cite{vafawitt}  and can also be motivated by requiring the cancellation  of
chiral anomalies in $p$-branes \cite{blumharvey}.  Here $C^{(3)}$ is the
Ramond--Ramond  three-form and $X_8$ is an eight-form made out of the
Riemann curvature.    The presence of this term in the effective  action, together with supersymmetry, must imply the presence of an infinite number of other terms.  Some of these  terms will be the subject of this article.

In the next section I will review  the evidence that certain `protected'  terms
in the low energy effective type II string action compactified on $S^1$ are
precisely determined by one-loop effects in eleven-dimensional supergravity
compactified on $T^2$.   For example  the one-loop eleven-dimensional
four-graviton amplitude  leads to  a  $R^4$ term in M-theory \footnote{The
notation $R^4$   represents a particular  contraction of four Riemann curvature
tensors that will be reviewed below.} which is related by supersymmetry to the
$C^{(3)} \wedge  X_8$ term.    This term can be expressed in the type IIB
coordinates by making use of the equivalence of M-theory compactified on $T^2$
with  type IIB string theory compactified on $S^1$ \cite{aspinwall,schwarz}.
This leads to a term of the form  $f(\rho,\bar \rho) R^4$  term in the   type
IIB action  \cite{ggv} where $f$ is a $SL(2,Z)$-invariant  function of the
complex  IIB scalar field, $\rho$.  The modular properties of the IIB theory
are inhereted from the geometric $SL(2,Z)$ of $T^2$ in M-theory with $\rho$
identified with the complex structure, $\Omega$, of the torus.
The interplay between the duality symmetries of nine-dimensional type II string
theories and the  one-loop eleven-dimensional supergravity amplitude is 
encoded in $f$  which contains both perturbative terms  and   non-perturbative
D-instanton contributions.   Furthermore,   even though the  eleven-dimensional
loop amplitude  is  cubically divergent    the consistent identification of the
compactified theory with the IIB   theory determines a
specific finite renormalized value  for the  coefficient of the $R^4$ term in
eleven dimensions    --- this finite value would necessarily arise as the
regularized value  in any microscopic theory.    In fact, since the ultraviolet
divergence is proportional  to the volume of the compactification torus,  it
disappears in the zero-volume limit, and the exact
(non-perturbative) $R^4$ term of ten-dimensional type IIB \cite{gg} is
reproduced by  finite terms alone.

The  expression for the modular function $f(\rho,\bar \rho)$  in the IIB theory
will be motivated  in section 2.1  by matching  the known $R^4$  terms that
arise at tree level and one loop  in string perturbation theory together with
the expected structure of  multiply-charged D-instanton contributions.   A single   D-instanton of charge $N= \hat m n$ can be identified, following T-duality, with configurations in the IIA theory in which the world-line of a  charge-$n$  D-particle winds $\hat m$ times  around a compact dimension.
Precisely the same expression will be obtained in section 2.2 from the low
momentum limit of the  one-loop four graviton scattering   amplitude in
eleven-dimensional supergravity compactified on $T^2$.   I will also make some
comments concerning the relation of the higher-order terms in the momentum
expansion of the one-loop supergravity amplitude with string loop diagrams
(related comments are made in \cite{russt}).

The  $R^4$ term  is related by supersymmetry to a large number of other terms.
In the language of the type IIB theory these terms include  a sixteen-fermion
term, $f_{16}(\rho,\bar \rho) \lambda^{16}$, where $\lambda$ is the complex
spin-$\half$ chiral fermion \cite{ggk}.    This is the analogue of the 't Hooft
fermion vertex  in a conventional Yang--Mills instanton background.   These
interaction terms can be expressed  as integrals over half the ten-dimensional
 on-shell  superspace (there is no  ten-dimensional off-shell superspace
formalism).  Such  protected terms again only receive perturbative
contributions at tree-level and at one loop in string theory and appear to be
determined entirely by one-loop diagrams in eleven-dimensional supergravity.
A brief overview of  such terms is given in section 2.3.

The expressions for the instanton contributions in the expansions of these interaction terms defines a measure on the space of  charge-$N$ D-instantons.   In \cite{gg2} it was argued that this should  equal the partition function of the zero-dimensional  Yang--Mills   model, which is an integral over bosonic and fermionic matrices in the Lie algebra of $SU(N)$ \cite{wittd}.    This, in turn,  is related to the bulk term in the Witten index for charge-$N$ D-particles \cite{sethi,yi}.  This circle of arguments which is presented in detail in \cite{gg2} will be briefly reviewed at the end of section 2.3.

It would be very interesting to discover how much of the structure of the
string perturbation expansion  can be determined from   eleven-dimensional
supergravity beyond these protected terms.  For example, the $R^4$ term is the
lowest-order term in a momentum expansion of   the exact  four-graviton
scattering amplitude in type IIB string theory.   There are very likely to be terms of higher order in momenta that are also protected by supersymmetry.    To what extent can the
  momentum expansion of the effective action be determined by
perturbation theory in eleven dimensions?
A more limited question is to what extent the momentum expansion of the  string
tree-level amplitude  can be reproduced by eleven-dimensional supergravity
perturbation theory?   This depends on the systematics of the multi-loop
diagrams in eleven-dimensional supergravity as will be described in  section 3.  
These diagrams have finite pieces that depend on the moduli of the compact
dimensions.  New primitive divergences  arise corresponding to terms which have
derivatives acting on $R^4$,  denoted symbolically as  $\partial^{2n}  R^4$.
It would be interesting to see whether these might be determined by the same kind of arguments that led
 at one loop  to the determination of the $R^4$ term.   However,  the arguments
that  I  present   beyond  one loop are based on dimensional analysis rather than
explicit evaluation of the rather complicated Feynman diagrams.  In this way it is easy to see how  specific multi-loop diagrams could give specific finite
terms in the M-theory action that correspond to terms that arise from string perturbation theory.   However, it seems likely that the values of the  counterterms are fixed  unambiguously only at low orders in the momentum expansion  where they are protected by supersymmetry.

\section{Higher-order terms in M-theory}

\subsection{\bf $SL(2,Z)$-invariant $R^4$ terms in type IIB}

The leading perturbative  contributions  to the   four graviton scattering
amplitude in  type IIB superstring theory are of the form (in string frame)
\be\label{leading}
S_{R^4}^{pert} = \int \sqrt{g^B} \left(\zeta(3) \rho_2^2  + {\pi^2 \over 3}
\right)  t^8t^8 R^4,
\ee
where  $g_{\mu\nu}^B$ is the IIB string-frame metric and
\be\label{rdef}
\Rfour \equiv t^{\mu_1\dots \mu_8} t_{\nu_1\dots \nu_8}
R^{\nu_1\nu_2}_{\mu_1\mu_2} \cdots R^{\nu_7\nu_8}_{\mu_7\mu_8},
\ee
where the rank-eight tensor, $t^8$, is defined in  \cite{greenschwarz}.  The
first term in (\ref{leading}) is the tree-level contribution
\cite{grisaru,grosswitt}  that is of order $\alpha^{\prime 3}$ relative to the
leading Einstein--Hilbert term.  The second term is the one-loop contribution
\cite{greenschwarzclosed}  and has no dependence on the dilaton in the string
frame.   Non-perturbative contributions to the $\Rfour$  term  also arise from
single D-instantons with charge $N$    \cite{gg} which give  an infinite series
of non-perturbative contributions   that have the form
\be
S_{R^4}^{nonpert} =  \sum_{N>0} c_N(\rho_2) (e^{2\pi i N\rho} + e^{-2\pi i N
\bar \rho}) \Rfour.  \label{dterms}
\ee
 One way of counting these D-instantons makes use of T-duality  between the IIB
and the IIA theories in nine dimensions.   From the IIA point of view a
D-instanton  is associated with the world-line of a  D-particle of charge $n$
and mass $e^{-\phi^A} n$   (where $\phi^A$ is the IIA dilaton)  winding  around
the compact ninth dimension of radius $r_A$.   The euclidean action of this
configuration is   $S= 2\pi \hat mn  (C^{(1)} + i r_A e^{\phi^A})$ where $C^{(1)}$
is the IIA one-form and $\hat m$ is the winding  number of the world-line.     After
T-duality this leads to the expression $S= 2\pi N \rho$ for the D-instanton
action where $N=\hat mn$.    Although the coefficients $c_N$ are probably very hard
to determine directly they are fixed by the requirement that the total action
should be invariant under $SL(2,Z)$ transformations.   This means that it must
have the form,
\be 
 S_{R^4} \equiv S_{R^4}^{pert}  + S_{R^4}^{nonpert}   =  \int \sqrt{g^B} \rho_2^{1/2}\, f(\rho,\bar \rho) \Rfour\,  d^{10}x
,\label{rtot}
\ee 
where  $f(\rho,\bar \rho)$  is  a modular function since $R$ is invariant under
$SL(2,Z)$ transformations in the Einstein frame.

The precise coefficients of the  known perturbative contributions,  $S^{pert}$,
together with the general form of the instanton corrections motivates  the
suggestion  \cite{gg} that $f$ is given by
\be 
f(\rho,\bar \rho) = \pi^{-2}  \sum_{(\hat l_1,\hat l_2)\ne (0,0)} { \rho_2^{3/2} \over
|\hat l_1 +
\rho \hat l_2|^{3}} 
=  \pi^{-2} \zeta(3)E_{3\over 2}(\rho,\bar \rho), \label{fdeff}
\ee 
where  the nonholomorphic Eisenstein series $E_s$ is  defined by \cite{terras}
\bes 
  \zeta(2s)E_s(\rho,\bar \rho)& =&
  \sum_{(\hat l_1,\hat l_2)\ne (0,0)} { \rho_2^s \over |\hat l_1 +
\rho \hat l_2|^{2s}}   \nn\\
 & =&     \zeta(2s)
\rho_2^{s} + {\sqrt \pi \Gamma(s-\half) \zeta(2s-1) \over
\Gamma(s)}\rho_2^{1-s} + {\calR}_s.\label{nonholo}
\ees
The last equality is a large-$\rho_2$ expansion  and
${\cal R}_s$ indicates a specific  sum of exponentially decreasing terms.  In
the special case $s=3/2$ this sum is given by
\be 
\calR_{{3\over 2}} =  \sum_{N >0}\left( \sum_{N|\hat m} {1\over {\hat m}^2}\right) N^\half
 \left(e^{2\pi i N \rho} +  e^{-2\pi i N\bar
\rho}\right) 
 \times \sum_{k=0}^\infty
(2\pi N\rho_2)^{-k}{\Gamma(\half -k) \over
\Gamma(-\half -k)}, 
\label{nonperty}
\ee 
where $\sum_{N|\hat m}$ indicates a sum over the divisors of $N$.
There are only two power-behaved terms in  the expansion (\ref{nonholo})  and
they  correspond precisely
to the known  tree-level and one-loop terms in the $R^4$ effective action of
the
  IIB   theory, while the series of exponentials in (\ref{nonperty})
correspond to D-instanton corrections with the expected  instanton number
$N=\hat mn$.   This lends weight to the
suggestion that (\ref{fdeff})
 is the exact result, in which case  there should be
a perturbative non-renormalization theorem \cite{gg} that forbids
corrections to the
 $R^4$ term beyond one loop.    Recently there has been a certain
amount of evidence for the validity of  such a theorem \cite{anton,berko}
(although  an apparent  contradiction in the literature \cite{jengo} deserves
closer analysis).
The measure factor $\sum_{N|n} {1\over n^2} $ in the instanton sum
(\ref{nonperty})  can  be related \cite{gg2} to the expression for the Witten
index   of relevance in the analysis of D-particle threshold bound
states   \cite{sethi}\cite{yi} as will be described at the end of section 2.3.

\subsection{\bf One loop in eleven dimensions}

The leading low energy behaviour of the one-loop four-graviton amplitude of
eleven-dimensional supergravity compactified on a torus with radii
$R_{10}$ and $R_{11}$   was considered in \cite{ggv}.     Here we
will consider  the complete momentum dependence of the same amplitude (a
similar argument was also given in \cite{russt}) and expand the expression in a
power series in the Mandelstam invariants,
$S,T$ and $U$,
\be 
S  = -(k_1+k_2)^2  , \qquad T= -(k_1 + k_4)^2,  
  \qquad U = -(k_1 + k_3)^2,
\label{mandelinvs}
\ee 
so that  S+T+U =0.

The terms of interest arise from the   sum of all one-loop diagrams with four
external  gravitons and with the graviton,   gravitino or  antisymmetric
three-form potential circulating around the loop.
This sum   is most succinctly calculated in a first-quantized light-cone gauge
formalism in which  the amplitude is described as a trace over the states of an
eleven-dimensional super-particle circulating around the loop and  coupled to
the four external gravitons by vertex operators (as in \cite{ggv}).   The
result is given by
\be\label{compamp}
A_4 = {1\over \kappa_{11}^2}\tilde K[ I(S,T)  + I(S,U) + I(U,T)],
\ee
where $\tilde K$ is the linearized approximation to $R^4$
(which is eighth order in momenta and symmetric under the interchange of
any pair of gravitons) and  the function  $I(S,T)$  has  the form of a  Feynman integral for a massless scalar field theory.  It is given by
\be 
I(S,T) =
 {1 \over\pi^{5/2} \calV_2}\int \prod_{r=1}^4 d\tau_r \int d^9q 
 \ \sum_{\{l_1,l_2\}}
e^{- G^{IJ}l_{I}l_{J}  \tau -    \sum_{r=1}^4 p_r^2 \tau_r},\label{feynint}
\ee 
where $\tau = \sum_{r=1}^4 \tau_r$ and  $q_i$ ($i=1, \cdots,9$) is the
nine-dimensional loop momentum transverse
to $T^2$.   The parameters $\tau_i$ label the relative positions of the four
vertices.   The sum is over the Kaluza--Klein momenta ($l_1$ and $l_2$) in the two compact
dimensions
($I,J=1,2$)  and the momenta in the legs of the loop are given by
\be
\label{momdef}
p_r = q + \sum_{s=1}^r k_s,
\ee
where the external momenta, $k_r^\mu$ satisfy $k_r^2=0$ and $\sum_{r=1}^4  k_r
=0$.
 The particular kinematic configuration has been chosen  in which the external
momenta
 have zero components in the directions of the torus, i.e. $k_r^I =
0$.   The eleven-dimensional
coupling constant, $\kappa_{11}$, has dimension, $(length)^{9/2}$ and will be
set equal to 1 in most of the following.
The inverse metric on $T^2$ is defined by
\be\label{metdef}
G^{IJ} l_I l_J = {1 \over  {\cal V}_2  \Omega_2} |l_1 + \Omega l_2|^2,
\ee
where ${\cal V}_2$ is the volume of the torus  with
complex structure,
$\Omega = \Omega_1 + i \Omega_2$   (where $\Omega_1$ is an angular
parameter and $\Omega_2 = R_{10}/R_{11}$).
Setting all the $k_r=0$ in (\ref{feynint}) gives the lowest-order result  of
\cite{ggv}.
The full amplitude (\ref{compamp})  gives rise to terms in the M-theory
effective action compactified
on $T^2$ of the form
\be\label{effact}
S_4 = {1\over \kappa_{11}^2} \int \sqrt{G^{(9)}} \calV_2
h(\calV_2,\Omega;\partial^2)R^4 d^9 x.
\ee
The function $h$ is a
modular function of $\Omega$ and its argument $\partial^2$
symbolically indicates derivatives acting on the fields in $R^4$,
corresponding to the dependence of $A_4$ on the momenta $k_r$.

After  completing a square in the exponent of (\ref{feynint}) and then
performing the shifted loop integral  the expression becomes
\be
I(S,T) = {\pi^2 \over \calV_2} \int    \prod_{r=1}^4 d\tau_r
\tau^{-9/2}  \sum_{\{ l_1, l_2\}}
e^{- G^{IJ} l_{I}  l_{J}  \tau   +{1 \over \tau} (S \tau_1\tau_3 + T
\tau_2\tau_4 ) }.\label{moreexp}
\ee 
This integral is to be evaluated in  the  region    $S, T<0$ where it converges  and then analytically  continued  to the physical region.

The  momentum-independent 
terms in (\ref{moreexp}) can be isolated  by writing
\be\label{momindef}
I(S,T) = I_0 + I'(S,T),
\ee
where
\be 
I_0  \equiv  I(0,0)   =   {\pi^{3/2}\over \calV_2}\int_0^\infty    d\tau
\tau^{-3/2} \sum_{\{ l_1, l_2\}}
e^{-\pi G^{IJ} l_{I}  l_{J}  \tau},\label{izerodef}
\ee 
which is the expression considered in \cite{ggv}.  It has a
divergence in the limit $\tau = 0$.  A double Poisson resummation reexpresses
$I_0$ as a sum over  $\hat l_1$ and $\hat l_2$ that may be identified with the winding numbers of the euclidean
world-line of the super-graviton around the directions $R_{11}$ and $R_9$ of 
 the torus, respectively.  The result is 
\bes
I_0 &=& \pi^{3/2} \int_0^\infty  d\tau \tau^{-5/2} \sum_{\{\hl_1,\hl_2\}}
e^{-\pi G_{IJ}\hl_{I}\hl_{J} {1 \over \tau}}\nn\\
&=& \pi^{3/2} \int_0^\infty d\hat \tau\hat\tau^\half  \sum_{\{\hl_1,\hl_2\}}
e^{- \pi G_{IJ}\hl_{I}\hl_{J}  \hat\tau}
,\label{windmod}
\ees
where $\hat \tau = 1/\tau$.  
This isolates the divergence in the  zero winding term ($\hl_1=\hl_2 = 0$).
This is presumably regularized by a microscopic theory, such as
the matrix model \cite{bfss}, but its regularized  value is also determined
uniquely by requiring that $I_0$ reproduce  the IIA and IIB string
theory $R^4$ terms in nine dimensions.  Since the loop diverges as
$\Lambda^3$, where $\Lambda$ is a momentum cut-off, the regularized
value of this term has the form $c \kappa_{11}^{-2/9}$ where $c$ is a
dimensionless constant.
 The remaining terms in (\ref{windmod})   depend
on the volume and complex structure of $T^2$  and are finite.  The $\hat \tau $ integral is trivial for these terms and  
(\ref{windmod})
 can be written as
\be\label{windmodagain}
I_0 =  c \kappa_{11}^{-2/9}  + \calV_2^{-3/2}
\zeta(3)E_{3\over 2}(\Omega,\bar \Omega).
\ee
Substituting  into (\ref{momindef}) and (\ref{compamp}) leads to  a $R^4$ contribution to the M-theory
effective action that can be expressed in terms of the IIB theory compactified
on a circle.\footnote{This makes use of the usual   relations between the
parameters of  M-theory on $T^2$  and the  type II string theories on $S^1$:
$r^A = (r^B)^{-1} = R_{10} (R_{11})^\half$, $
e^{\phi^B} =  (r^A)^{-1} e^{\phi^A}  = R_{11} / R_{10 }$,
where $r^A$ is the radius of the IIA circle in string units and   $\phi^A$ and $\phi^B$ are the dilatons of the two type II theories that are related by T-duality.}
  
 In the limit $\calV_2 \to 0$ the radius, $r_B$,  of the  IIB
circle becomes infinite and the second term in (\ref{windmodagain})
 dominates, leading to an expression for the action for the decompactified IIB
theory which coincides with (\ref{rtot}) (with $f$ defined by
(\ref{fdeff})).  The IIB string   tree-level contributions arise from terms in (\ref{windmod}) with $\hat l_1 \ne 0$ and  $\hat l_2 =0$ so the loop has non-zero winding only in the eleventh dimension.  The string one-loop
and D-instanton terms in  (\ref{rtot})   can be extracted from the $\hat l_2 \equiv \hat m \ne 0$ terms in   (\ref{windmod}) by   performing a Poisson summation that takes the winding numbers $\hat l_1$ into   Kaluza--Klein charges $n$ and identifying  $N= \hat m n$.     The $n=0$ term gives rise to the term in (\ref{rtot}) that is independent of the dilaton (the one-loop contribution) while the charge-$N$ D-instanton contributions   in  (\ref{nonperty})  come  from the $N\ne 0$ contributions.

Although the  zero winding number term with coefficient $c$  does not contribute in
the $\calV_2 \to 0$ limit it does contribute to the finite-$\calV_2$ amplitude.
By transforming to the type IIA
coordinates it is easy to see that $c$ is the coefficient of 
the one-loop contribution to the  $R^4$ term in the IIA theory in ten dimensions and
takes precisely the same value as the  one-loop term in $S^{pert}$, thereby
ensuring the T-duality of the type IIA and IIB theories \cite{ggv}.

There is an intriguing analogy between this calculation and the calculation of the exact prepotential of    four-dimensional $N=2$ supersymmetric Yang--Mills theory  from a one-loop amplitude in five dimensions \cite{launek}.

The momentum-dependent terms in $I(S,T)$  in  (\ref{momindef}) are contained in $I'(S,T)$ .  This is evaluated by separating the $l_1=l_2=0$ term from
the rest,
\be\label{iprimedef}
I'(S,T) = I^0(S,T) + \sum_{n=2}^\infty I_n(S,T).
\ee
The term with zero Kaluza--Klein momenta, $I^0$, gives a   contribution to the
amplitude that is non-analytic in the Mandelstam invariants while  $I_n$ is a
homogeneous  polynomial  in $S$ and $T$ of degree $n$.    After translating to
 string coordinates the term $I^0$ can be identified  with a  corresponding
one-loop term in string perturbation theory.   The polynomials $I_n(S,T)$,
$I_n(T,U)$ and $I_n(U,S)$ can be identified with terms that  should  arise from
multi-loop amplitudes in the supergravity theory.  More details are given in the Appendix  which overlaps with   \cite{russt}. 

 Compactification of the supergravity loop amplitude  to eight dimensions on
$T^3$ raises new issues associated with the presence of new instantons that are
identified with   wrappings of the   M-theory membrane world-volume around
$T^3$.  The complete expression for the $R^4$  term is now obtained by
combining the one-loop calculation with the constraints of U-duality
\cite{gv,ggv,kiritsis}.  Related issues arise in considerations of type I/heterotic duality in eight dimensions \cite{threshold,bachstrings}.
 Compactification to lower dimensions raises yet more
issues analogous to those that arise in the compactification of the matrix  model.   

 \subsection{\bf Sixteen-fermion and related terms}

The $f(\rho,\bar \rho) \Rfour$ term is one of a large number of  terms of the
same dimension that are related to each other by supersymmetry.   Prominent among these is the analogue of the 't Hooft
multi-fermion vertex of the IIB theory,
\be
\label{lamsixteenact}
S_{\lam16} = \int \sqrt{g^B} \rho_2^{1/2} f_{16}(\rho,\bar \rho) \lam16 d^{10} x,
\ee
 where $\lambda$ is the complex chiral  spin-$\half$ fermion.  This field transforms with weight
$3/2$ under the $U(1)$  denominator in the coset $SL(2,Z)/U(1)$
\cite{westschwarz,schwarz1,howewest} so that  a general $SL(2,Z)$
transformation acts as
\be
\label{modonlam}
\lambda \to    \left({c\bar \rho+d\over
c\rho+d}\right)^{3\over 4 }\lambda .\ee
As a  consequence of  $SL(2,Z)$ invariance of the action   it follows that  the
coefficient  $f_{16}$ in (\ref{lamsixteenact})   is a non-holomorphic   modular
form of weight $(12,  -12)$ (where the notation indicates the  holomorphic and
anti-holomorphic weights, respectively).

The expression for $f_{16}$ can also be obtained by a one-loop calculation in
eleven-dimensional supergravity compactified on $T^2$.  This time  the process
has  sixteen external gravitini in polarization states that correspond to
$\lambda$ in the IIB description. With this choice of  external states  the
loop diagram vanishes in the limit of eleven non-compact dimensions and is
finite for generic $T^2$.   It was evaluated in  a recent paper  \cite{ggk}
and the result is
\be 
 f_{16}  =
   {1 \over \pi^2 }   \rho_2^{3/2}  \Gamma (27/ 2)
         \sum_{( \hat l_1, \hat l_2)\ne (0,0)} {( \hat l_1 +  \hat l_2\bar \rho)^{24} \over
| \hat l_1 +  \hat l_2 \rho|^{27}}, 
 \label{Iresult}
\ee 
where $\hat l_1$ and $\hat l_2$ are the winding numbers of the world-line around the cycles of the torus.
This expression   has the large  $\rho_2$ expansion,
\bes
 \rho_2^{1/2} f_{16} & =&  \pi^{-2}\Gamma(27/2)\zeta(3) (\rho_2)^2
  +  \pi^{-2} \Gamma(23/2)  \zeta(2)    \nn\\
&& + 2^{24}
\pi^{23/2} \sum_{N>0}\left(\sum_{N|\hat m} {1\over {\hat m}^2}\right)
(N\rho_2)^{25/2}   e^{2\pi i   N \rho } (1 + O(\rho_2^{-1})),  \label{largef16}
\ees
which again indicates the presence of  string perturbation theory contributions
at tree-level and one-loop together with an infinite number of D-instanton
contributions (the anti D-instantons enter with coefficients that are  of
higher order in $\rho_2^{-1}$).  There are  presently  no sixteen-fermion perturbative 
string  calculations   in the literature with which to compare the  two
leading  terms.

In \cite{ggk} it was also shown that the functions $f_{16}$ and $f$ are related
by
\be
\label{covrel}
f_{16} (\rho,\bar \rho)  =     \rho_2^{12}{\cal D}^{12}f(\rho,\bar \rho).
\ee
The covariant derivative ${\cal D}$ is defined by
\be 
F_{d+2, \bar d}  =  {\cal D}_d F_{d,\bar d} 
=  i \left( { \partial\over
\partial
\rho}  + {d
\over (
\rho - \bar \rho)}\right) F_{d,\bar d},\label{covder}
\ee
where $F_{d, d'} $ is a    modular form  of  holomorphic weight $d$ and
anti-holomorphic weight $d'$ and
\be
\label{seqd}
{\cal D}^k F_{d,\bar d} \equiv {\cal D}_{d + 2(k-1)} {\cal D}_{d+ 2(k-2)}
\cdots
{\cal D}_d
F_{d,\bar d}.
\ee

The relationship (\ref{covrel}) between $f$ and $f_{16}$ is  consistent with
the consequences of linearized supersymmetry.  The physical on-shell fields of
linearized  IIB supergravity are contained in a superfield $\Phi (x,\theta)$
(where $\theta$ is a complex chiral Grassmann spinor of $SO(9,1)$) satisfying
the
holomorphic constraint $ D^* \Phi=0$ and the on-shell condition
$D^4    \Phi
=  D^{*4} \Phi^*$ where $D$ and $D^*$ are supercovariant derivatives
\cite{howewest} .   This field has an expansion in powers of $\theta$,
\be 
\Phi = \rho_0 +  a  - {2i \over g}\bar\theta \lambda
  - \cdots     - {i \over 48
g}\bar\theta\Gamma^{\mu\nu\eta}\theta\bar\theta\Gamma_{\eta}^{\
\sigma\tau}\theta
R_{\mu\nu\sigma\tau}+
\cdots,\label{phidef2}
\ee 
 where $\rho_0= \rho -a$ is a constant  background that defines the coupling
constant in the linearized theory,  $R$ is the linearized approximation to the
curvature  and $\cdots$ indicates the other terms in the expansion. The
matrices  $\Gamma^\mu$ are   $SO(9,1)$ gamma matrices and $\bar \theta =
\Gamma^0 \theta$ (with no complex conjugation).
 The linearized
supersymmetric on-shell action  has the form
\be\label{actdef}
S' =   {\rm Re } \int d^{10}x  d^{16} \theta \,  g^4  F[\Phi],
\ee
and the  various  component interactions
are contained  in   the $\theta^{16}$ term in the expansion of  $F[\Phi]$.
In addition to the $R^4 $ and $\lam16$  terms there are many others   --
some of these are singlets under the $U(1)$ (and arose at one string loop in
\cite{grosssloane}) while others carry a net even number of units of $U(1)$ charge \cite{ggk}  (some of these terms have also been discussed in \cite{partouche}).

Although the fully nonlinear theory cannot easily be expressed in terms of $\Phi$ the
leading behaviour of the charge-$N$ instanton term is obtained by substituting
$F =N^{-7/2} e^{2\pi i N \Phi}  \sum_{N|n} {1\over \hat m^2} $ into (\ref{actdef}).
In this manner the leading instanton contributions to the  $R^4$, $\lam16$
and related terms are  all contained in  a simple superspace expression.

The expression for the measure on single charge-$N$ D-instantons,
$\sum_{N|n} {1\over \hat m^2}$, 
 can be read off from the expressions for the instanton contributions to the various processes described above.   This can then be compared with the expression that should follow from the matrix description of D-instantons. According to \cite{wittd} this is determined by the  partition function for $U(N)$ 
ten-dimensional supersymmetric Yang--Mills theory reduced to zero dimensions, in which case the configuration space consists of the ten $N \times N$ bosonic matrices, $A_\mu$, and sixteen $N\times N$ fermionic matrices, $\psi^a$.  The partition function is given by
\begin{equation}
\label{instmeasure}
Z_{U(N)} = \int d^{10} y d^{16} \epsilon Z_{SU(N)},
\end{equation}
where the center of mass super-translation zero modes have been separated from the measure factor,
\begin{equation}
\label{measdef}
Z_{SU(N)} = \int_{SU(N)} DA D\psi \exp(-S(A,\psi)).
\end{equation}
This integral, which enters into the matrix model of \cite{matrixmod},  has not been explicitly evaluated apart from the $N=2$ case considered in \cite{sethi}\cite{yi} (which gave $Z_{SU(2)} = 5/4$).  However, equating it with the  measure for $N$ coincident D-instantons obtained from the $\Rfour$ and related terms strongly suggests that 
\begin{equation}
\label{measure}
Z_{SU(N)} = \sum_{N|\hat m} {1\over {\hat m}^2}, 
\end{equation}
for all $N$ \cite{gg2}.  It would obviously be of interest to check this by explicit evaluation of the finite-dimensional  integral (\ref{measdef}).

\section{{\bf Tree-level string theory and  multi-loop supergravity.}}

We have seen that the lowest-dimensional terms in the string  theory effective action correspond to terms that arise at one loop in eleven-dimensional supergravity.  We would now  like to see whether   higher-order string-theory terms can be obtained from multi-loop amplitudes in eleven dimensions.  To be explicit we  will 
 compare the expansion of the ten-dimensional IIA  string tree
amplitude with the compactification of multi-loop diagrams of
eleven-dimensional supergravity on a circle of radius $R_{11}$.
The   calculation of multi-loop diagrams is obviously very
difficult but we will be able to infer  certain systematic properties from dimensional considerations.

First, consider the tree amplitude for four-graviton scattering
in either of the type
II superstring theories which is given by
\bes
A_4^{tree} &=& e^{-2\phi}{\tilde K\over stu}
{\Gamma(1-\alpha' s)\Gamma(1-\alpha' t)\Gamma(1-\alpha' u) \over
\Gamma(1+ \alpha' s)\Gamma(1 + \alpha' t)\Gamma(1  + \alpha' u)} \nonumber\\
&=& e^{-2\phi} {\tilde K  \over stu} \exp\left(\sum_{n=1}^\infty
{2 \zeta(2n+1) \over 2n+1} (s^{2n+1} + t^{2n+1} +
u^{2n+1})\right),
\label{treeamps}
\ees
where the string-frame Mandelstam invariants are related to those in the
M-theory coordinates by
\be\label{mandelstring}
s =  {S\over  R_{11}}, \qquad t =  { T\over  R_{11}}, \qquad u = {U \over  R_{11}}.
\ee
It is important that every term in the exponent can be expressed as a
polynomial in $s$ and $t$ multiplied by $stu$, as can be seen from
the identity,
\be 
 s^{2n+1} + t^{2n+1} + u^{2n+1} = stu\left[
\sum_{r=1}^n \sum_{q=0}^{2n-2r} \right.    {(2n+1)! \over r!
(2n+1-r)!}   (-1)^q s^{2n-1-r-q} t^{r+q-1}   \bigg],
\label{factout}
\ee 
where $n\ge 1$.  This means that the massless poles only contribute to
the first term in the expansion of the exponential.

When expressed in terms of the Mandelstam invariants in the M-theory metric
the  expression (\ref{treeamps})  has the low-energy expansion,
\bes
&&A_4^{tree}  \sim   \tilde K  \left( {1\over STU}  +
{2\zeta(3)\over R_{11}^{3}}   +{ 2 \zeta(5)\over R_{11}^{5}}(S^2+ST+T^2)
+ {2 \zeta(3)^2\over  R_{11}^{6}}STU    \right. \nonumber\\
&&\left. +
{2\zeta(7)\over R_{11}^{7}}(S^4  + 2S^3T +3 S^2T^2 +2 S T^3 + T^4)
  + { 2 \zeta(3)
\zeta(5)\over R_{11}^8} STU(S^2+ST + T^2) \right. \nonumber\\
&& \left. + {2\zeta(9)\over R_{11}^9}(S^6 + 4 S^5T + \cdots + T^6)   +
{4\over 3 R_{11}^9}\zeta(3)^3 S^2T^2U^2 + \cdots \right).
\label{lowtree}
\ees
The first term in the expansion combines with the kinematic factor,
$\tilde K$, to give the tree-level  amplitude that is
described by  the Einstein--Hilbert action of eleven-dimensional
supergravity compactified on a circle of radius $R_{11} =
e^{2\phi/3}$.
The subsequent term, with
coefficient $\zeta(3)$  is the  term considered earlier  that  comes from
the   $R^4$ term in the effective action.    We saw that this term is
reproduced by the one-loop diagram of
eleven-dimensional supergravity compactified on a circle.
The higher-order terms in  (\ref{lowtree}) come from terms in the effective
action with
derivatives acting on $R^4$.    The question is whether the whole
series might be reproduced by summing loop diagrams of
eleven-dimensional supergravity compactified on a circle.

\begin{figure}[htb]
\vspace{9pt}
\centerline{ {\epsfbox{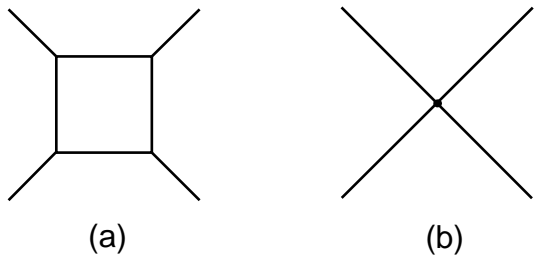}}}
\caption{(a)  The finite part of the one-loop diagram contributing to the $R^4$
term of compactified eleven-dimensional supergravity comes from the non-zero
windings of  the circulating particles.   (b)  The primitively divergent  $R^4$
term that arises from the regularized zero winding sector is regularized to a
specific finite value consistent with supersymmetry and various duality
symmetries.}
\label{fig:fone}
\end{figure}

 We will restrict ourselves here
to considering some simple multi-loop
 contributions that should reproduce the first few  terms in the low-energy
expansion of the string tree-level amplitude, (\ref{lowtree}).  We have already
seen that  the constant term proportional to
$\zeta(3) (R_{11})^{-3}$  comes from the finite part of the
one-loop diagram of figure 1(a).
Figure 1(b)  represents the counterterm for the
 primitive divergence of the one-loop diagram that is regularized to a
finite value by the considerations outlined earlier.

\begin{figure*}[htb]
\vspace{9pt}
 \centerline{{\epsfbox{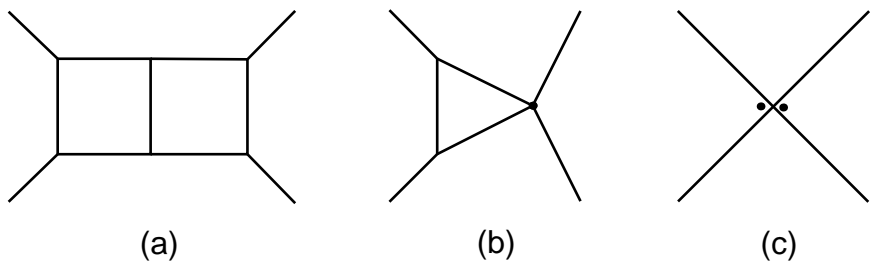}}}
\caption{(a) A representation of the non-zero windings of the particles
circulating in a two-loop   contribution to four-graviton
scattering.   The  dimension, $(momentum)^{20}$, of this finite contribution
could  give terms of the form $s^3$ in the ten-dimensional string  tree-level
amplitude.  (b)  A contribution in which one loop is replaced
by the $R^4$ counterterm.  has dimension $(momentum)^{17}$ and could contribute
terms of the form $s^2$ to the string tree-level amplitude.
(c)  A counterterm for the two-loop primitive divergence.}
\label{fig:ftwo}
\end{figure*}

Just as with the one-loop diagrams, there are  finite
contributions from multi-loop diagrams
which come from non-zero
windings of the internal loops around the compact dimension(s).
At each order there is also a new primitive divergence.
For example, there are many two-loop diagrams that
must be summed to give the complete amplitude.   These
have  superficial degrees of divergence $\Lambda^{20}$, where $\Lambda$
is a momentum space cut-off and is $O(\kappa_{11})^4$.
However, we are discussing terms in the amplitude  proportional to derivatives
acting on  $R^4$  so   there are at
 least eight powers of the external momenta, reducing the naive
divergence to $\Lambda^{12}$, or less (depending on the number of derivatives).
 This leads to new (two-loop)
primitive divergences indicated by figure 2(c), where the two dots indicate the
fact that this is a combination of  two-loop counterterms.
More generally, at $n$ loops there are  primitive divergences of the form
$\Lambda^{9n-6}R^4$, where the powers of the cut-off $\Lambda$ may be
 substituted by powers of the external momenta.  The ultra-violet
divergences coming from
the zero winding-number sector give rise to counterterms
that are independent of $R_{11}$.

 However, just as with the one-loop diagrams there will
be finite terms arising from the effects of the internal propagators
winding around the compact direction(s).
These are
collectively represented, in the case of the two-loop diagrams,
 by the ladder diagram in figure 2(a) which has
dimension $(momentum)^{20}$.
 After accounting for the eight powers of the
external momenta in the overall $R^4$, twelve powers of momenta remain
that must be replaced by
appropriate powers of the dimensional parameters, $R_{11}$ and
$S,T,U$.  This diagram may give a contribution  to the tree-level
string amplitude which is proportional to $(R_{11})^{-3}$.   In such a
contribution there are
 nine powers of momentum left to be soaked up by a combination involving
Mandelstam invariants.   In the analogous
expressions derived from string theory, such as (\ref{lowtree}),
each Mandelstam invariant  comes along with one power of
 $R_{11}^{-1}$.   Thus, the  nine
powers of momenta  should be associated with a
linear combination of
$S^3/R_{11}^3$, $T^3/R_{11}^3$, $S^2T/R_{11}^3$ and $T^2 S/R_{11}^3$.  We would
like to identify this with the term in (\ref{lowtree}) that is cubic in the
invariants.
At  $n$ loops the  Feynman diagrams analogous to figure 2(a)
can  give finite contributions  of the form
$s^{3n-3}$ in the expansion of the string tree amplitude (\ref{lowtree}).

In addition to these finite contributions there are also   many other loop
diagrams in which the  counterterms are
inserted as vertices.  In fact, the
 diagram with the smallest primitive degree of divergence (other than the
one-loop diagrams of figure 1) is the one shown
in figure 2(b), in which the vertex indicated by a   dot
 is the $R^4$ counterterm that was
required to make sense of the one-loop diagram.   Since
all the particles of the supermultiplet
circulate around the loop the supersymmetric
partners of the $R^4$ vertices are also involved. These
couple the two external gravitons to two $C^{(3)}$'s,  or to 
two gravitini.  Figure 2(b) is
meant
to represent all diagrams of this order containing a single one-loop
counterterm.
These  have  dimension $(momentum)^{17}$ so that
after allowing for the eight powers of external momenta in $R^4$
 there are nine powers of momentum left to account for.   These
diagrams have not been explicitly evaluated but if it is assumed that they
contribute to the   string tree-level
process there must be  an overall
factor of $e^{-2\phi^A}= R_{11}^{-3}$, absorbing three of the nine
momentum powers.  Then dimensional counting
implies that the
the only possible finite function is {\it quadratic} in the invariants,
$s,t,u \sim S/R_{11}, T/R_{11}, U/R_{11}$.
This quadratic term ought to be precisely the first
momentum-dependent correction to the $R^4$ term deduced by expanding
the tree diagram.   It should therefore turn out that an explicit calculation
of figure 2(b) gives the finite contribution,
\be
\label{nameds}
{2\zeta(5) \over R_{11}^3} \tilde K \left({S^2 + ST + T^2 \over R_{11}^2}
\right),
\ee
where $\tilde K$  is the linearized approximation to $\Rfour$ and is eighth-order in the external momenta.    Although the  diagram in figure 2(b) has not been explicitly evaluated it  is tempting to imagine that it has the form  $\tilde K I(S,T,U)$, where $I$ is the expression for the loop amplitude of the same geometry in scalar field theory.  This  has a coefficient  that includes a factor of $\zeta(5)$  just as there was a  factor of $\zeta(3)$ in the finite part of the box diagram of figure 1(a).  Verifying that the  precise coefficient is the one in (\ref{nameds})  would lend support to the suggestion that  quantum corrections to
compactified
eleven-dimensional supergravity coincide with the string tree-level expressions at this order.   In a similar manner the term with coefficient $\zeta(7)$ in (\ref{lowtree}) could arise from the diagam with two $R^4$ vertices and two propagators.     However,  this exhausts the contributions from  one-loop diagrams with $R^4$ vertices and  it is not at all clear how the terms of higher order in momenta in (\ref{lowtree})  would emerge in any systematic manner. 
 
Of course, such  diagrams also contribute to string loop effects.  For example, there are  the generalizations of  the non-analytic  one-loop massless threshold term, (\ref{tenlog}).
In either of the type II string theories  these terms can be   written symbolically as 
\be 
 \int \sqrt{g_E^{(10)}}  s_E^n f_n(t_E/s_E) R^4 d^{10}x  = \int \sqrt{g^{(10)}} e^{{n-1\over 2}\phi^A} s^n f_n(t/s) R^4 d^{10}x,\label{thresh}
\ee 
where $g_E$ is the type II string metric in the Einstein frame.     The function  $f_n(t/s)$ involves up to $n$ powers of $t/s$ as well as factors of  $\log( t/(s+t))$ and $\log (s/(s+t))$.  The $n=1$ case is given implicitly by (\ref{tenlog}).  Since these  terms arise in string perturbation theory, which is an expansion in $e^{2\phi}$,  it must be that  $n =4m+1$, where $m$ is an integer.
In the Einstein frame these terms   are independent  of the dilaton so they are inert under the action of $SL(2,Z)$ in the IIB theory.   They are therefore not constrained by S-duality.  In
the IIA theory such terms are directly related to analogous thresholds in  genus-$(m +1)$ multi-loop   eleven-dimensional  supergravity amplitudes.   These Feynman diagrams are  dimensionally of the form $({\rm momentum})^{9m+3} R^4$.  After compactification their contribution  to the genus-$(m+1)$ string amplitude has a factor of $e^{2m \phi^A} = R_{11}^{3m}$ from   powers of the coupling.   The total momentum dimension of this contribution  is then written as  $R_{11}^{3m}
({\rm momentum})^{12m+3} R^4 $, which must therefore be made up of terms of the form    $e^{2 m  \phi^A } s^{4m+1} f_n(t/s) R^4$ in the string coordinates, using the fact that $s = S/R_{11}$.  This agrees in structure with the terms in (\ref{thresh}) if the function $f$ contains the appropriate logarithmic terms.

There
are   many complications in understanding in detail the systematics
of the
correspondence between the higher-loop supergravity diagrams and string
diagrams.    Whereas the $R^4$ and related terms of the same dimension are integrals over half the superspace,  terms with more derivatives are integrals over a higher fraction of the superspace.   Each power of momentum is equivalent to two powers of $\theta$ so that terms with   less than eight powers of momentum acting on $R^4$ should be protected and may be determined in this manner.   This would include the terms up to the $\zeta(3)^2$ term in    (\ref{lowtree}) and possibly also the $\zeta(7)$ term. 
Whether it is possible to go beyond this and relate  terms in string perturbation theory  at higher order in the momentum expansion   to eleven-dimensional supergravity is much less obvious.  

\vskip 0.5cm
\noindent{\it Acknowledgements}:

I am grateful to   Nati Seiberg,  Paul Howe and Adam Schwimmer, as well as  participants at the Aspen Center for Physics Workshop on String Dualities  in the Summer of 1997, for useful discussions.

\appendix\section{Momentum dependence of  loop amplitude}

The momentum dependence of the four-graviton amplitude is contained in the functions $ I_n(S,T)$ ($n=2, \cdots, \infty$)  in  (\ref{iprimedef}).  This  will be a homogeneous polynomial of degree $n$. There
is no $n=1$ term  after adding the
contributions of $I(S,U) $ and $I(T,U)$ since there is no nonvanishing  linear
symmetric combination.   The
term with zero Kaluza--Klein momenta is given by
\be  
  I^0(S,T)   =  {1 \over \calV_2}  \int\prod_{r=1}^4d\tau_r (\tau)^{-9/2} 
 \left(e^{ (S \tau_1\tau_3 + T \tau_2\tau_4 )/\tau } -1\right) ,\label{zerokkdef}
\ee 
where the $-1$ accounts for the fact that the constant terms are
contained in $I_0$.
The variables  $\tau_r$ may be redefined by 
\be 
\omega_1  ={\tau_1 \over \tau},\quad
 \omega_2 = {1 \over \tau} (\tau_1 + \tau_2)  , \quad
 \omega_3 ={1 \over \tau}  (\tau_1 + \tau_2 + \tau_3) , \quad \tau = \sum_{r=1}^4 \tau_r ,\label{varnew}
\ee 
which are to be integrated over the region   $0 \le \omega_1 \le \omega_2 \le \omega_3 \le 1$ so that 
\be 
I^0(S,T) =  {1 \over \calV_2}
\int_0^\infty  d\tau  \int \prod_{r=1}^3
d\omega_r (\tau)^{-3/2}
{\left(e^{-Q(S,T;\omega_r)\tau }-1\right)}, \label{newizerodef}
\ee
where
\be\label{qdef}
Q(S,T;\omega_r) = - S \omega_1 (\omega_3 -\omega_2) -  T (\omega_2-\omega_1) (1 -\omega_3).
\ee
The integral in (\ref{newizerodef}) is simply evaluated by noting that
$\partial I^0 /\partial Q = \sqrt \pi \int \prod d\omega_rQ^{-\half}$
so that 
\be 
I^0(S,T)  \equiv  2\sqrt \pi {\cal S}^{\half} =2 \sqrt \pi \int \prod d\omega_r
(-Q(S,T;\omega_r))^\half,\label{noname}
\ee
where, more generally, ${\cal S}^n$ is defined by
\be\label{gensdef}\calS^n = \int \prod d\omega_r
(-Q(S,T;\omega_r))^n.
\ee
Similarly, $\calT^n$ and $\calU^n$ will be defined by cyclically permuting
$S$, $T$ and $U$ in the function $Q$.  

This non-analytic term (\ref{noname}) in nine-dimensional M-theory
translates into a similar term in either nine-dimensional string
theory (as was also noticed in \cite{russt}) making
use of  the familiar relation, by $g^{(9)} = R_{11} G^{(9)}$,  between the  nine-dimensional 
 type II metric, $g^{(9)}_{\mu\nu}$,  and  the M-theory metric,   $G^{(9)}$,
compactified on $T^2$. 

Using the relation between the M-theory and string theory  Mandelstam invariants (\ref{mandelstring})  it follows that
\be 
 2\sqrt \pi\int \sqrt{G^{(9)}}R^4 (\calS^\half+
\calT^\half + \calU^\half) d^9x   
=
2\sqrt \pi\int \sqrt{g^{(9)}}
R^4(\cals^\half+\calt^\half + \calu^\half)d^9x,\label{masslessth}
\ee 
 where the expressions
$\cals,\calt$ and $\calu$ are defined in terms of the Mandelstam
invariants of string theory.
The analogous expression in the limit of decompactification, $R_{10}
\to \infty$, is of the form,
\bes
 &&\int \sqrt {G^{(10)}} R^4(
\calS\ln \calS + \calT\ln \calT + \calU \ln \calU)d^{10} x \nn\\
 &&\qquad\qquad =   \int \sqrt {g^{A(10)}} R^4(
\cals\ln \cals + \calt\ln \calt + \calu \ln \calu)d^{10} x.  
\label{tenlog}
\ees 
The scale of the logarithms cancels out after using the condition
$s+t+u=0$. Both  (\ref{masslessth}) and (\ref{tenlog}) have imaginary parts
corresponding to  the massless normal
thresholds determined by unitarity.  However, the real parts, which
might have given rise to arbitrary constants,  are here
fixed to precise values.

The   terms $I_n(S,T)$ in (\ref{iprimedef})  are given by by expanding the exponential in (\ref{moreexp}) giving,
\bes
I_n(S,T) &=&  \int_0^\infty  {d\tau \over  \tau^{{3\over 2}
-n} } \int \prod_{r=1}^3   d\omega_r 
  \sum_{( l_1, l_2) \ne (0,0)} 
e^{- G^{IJ} l_{I} l_{J}   \tau }{
(-Q(S,T;\omega_r))^n\over n!} \nonumber\\
& = & \Gamma(n-\half) \zeta(n-\half)  E_{n-\half}(\Omega,\bar \Omega)   \int \prod_{r=1}^3   d\omega_r 
{(-Q(S,T;\omega_r))^n\over n!} .\label{genterm}
\ees
Putting all the terms together the  complete expression for the $R^4$ term in the one-loop
effective action for eleven-dimensional supergravity on $T^2$ is 
given by (\ref{effact}) with the
function $h$ defined by the amplitude $A_4$ in (\ref{compamp}) where
\bes
&& I(S,T)+ I(T,U) + I(U,S) = \nonumber\\ 
&&I_0 +    2\sqrt \pi 
\calW^{\half}  +  \sum_{n=2}^\infty    {\calW^n \calV_2^{n-{3\over
2}} \over  n!}   \left[ \Gamma(n-\half)\zeta(2n-1)
\left({R_{10} \over R_{11}} \right)^{n-\half}\right. \nn\\
&&\left. +  \sqrt \pi
\Gamma(n-1) \zeta(2n-2) \left({R_{10} \over
R_{11}} \right)^{{3\over  2} - n}\right]  + {\rm non-perturbative \ terms} , 
\label{malll}
\ees
and
\be\label{calWdef}
\calW^n = \calS^n+\calT^n+\calU^n.
\ee
The $S$, $T$ or $U$-dependent  terms in (\ref{malll}) exponentiate in the form
$e^{S\calV_2}$ which vanishes in the infinite volume limit,
$\calV_2 \to \infty$.

It follows from (\ref{malll}) that the expression for the 
nine-dimensional effective action arising from the  one-loop 
supergravity 
amplitude can be written in IIA coordinates  as
\bes
&&  \int   \sqrt{g^{A(9)}}r^A R^4
\left[  I_0 +{1 \over r_A}  \calw^{\half}  +\sum_{n=2}^\infty 
  \left( {1 \over r_A^2} {r_A^{2n}
\calw^n \over n!} + e^{-2\phi^A}{ e^{2n
\phi^A}\calw^n \over n!}\right) \right]d^9x    \nn\\
&&\qquad\qquad\qquad\qquad\qquad +  {\rm non-perturbative \ terms}, 
\label{iiares} 
\ees
where
\be\label{calwdef}
\calw^n = \cals^n + \calt^n + \calu^n,
\ee
and $I_0$ can be reexpressed in type IIA parameters, $r_A$ and $\phi^A$.
The first sum  in round parentheses
 gives terms of the form  $e^{r_A^2 s}$  which are 
exponentially suppressed in the ten-dimensional limit, $r_A \to \infty
$.   The ten-dimensional limit in the IIA theory is therefore given 
by a power series in
$e^{2\phi^A}s$ or, for constant dilaton, by a series of powers of  $g_A^2 s$ where 
$g_A = e^{\phi^A}$.  Each power, $n$,  should presumably be identified with
a contribution from a string loop amplitude of genus $n+1$.  This
suggests that the leading low-momentum term at  genus $n$  of the IIA
theory should be of order $(g_A^2 k^2)^{n-1}$, where $g_A =
e^{\phi^A}$ for constant $\phi^A$ (and $k^2$ represents some
combination of $s,t$ and $u$).

In the IIB coordinates the action becomes 
\bes
&&\sum_n \int   \sqrt{g^{B(9)}}  r^B R^4  \left[  I_0 + {1 \over r^B} \calw^{\half}   +\sum_{n=2}^\infty 
\left({1 \over n!} {\calw^n \over r_B^{2n}}  +
{e^{-2\phi^B}\over
n!} {\calw^n e^{2n\phi^B} \over r_B^{2n}}\right)\right]d^9x \nonumber\\
  &&\qquad\qquad\qquad\qquad\qquad + {\rm non-perturbative \ terms} .
\label{iibres}
\ees
All the momentum-dependent terms in this expression
vanish in the limit $r_B\to
\infty$ with fixed coupling $e^{\phi^B}$.   This  means that these terms do not contribute to
the loop amplitudes of the ten-dimensional type IIB theory.   However, the  four-graviton loop amplitudes are identical in the IIA and IIB  theories as a consequence of T-duality.  This means that  
there must be  terms in an expansion in powers of $ e^{2\phi^B} s =   g_B^2 s $ that have another eleven-dimensional origin.   For example,  such terms can arise from eleven-dimensional multi-loop diagrams compactified on a torus.

\end{document}